# Towards Reliable and Explainable AI Model for Solid Pulmonary Nodule Diagnosis

Chenglong Wang, Yun Liu, Fen Wang, Chengxiu Zhang, Yida Wang, Mei Yuan, Guang Yang

*Abstract* — Lung cancer has the highest mortality rate of deadly cancers in the world. Early detection is essential to treatment of lung cancer. However, detection and accurate diagnosis of pulmonary nodules depend heavily on the experiences of radiologists and can be a heavy workload for them. Computer-aided diagnosis (CAD) systems have been developed to assist radiologists in nodule detection and diagnosis, greatly easing the workload while increasing diagnosis accuracy. Recent development of deep learning, greatly improved the performance of CAD systems. However, lack of model reliability and interpretability remains a major obstacle for its large-scale clinical application. In this work, we proposed a multi-task explainable deep-learning model for pulmonary nodule diagnosis. Our neural model can not only predict lesion malignancy but also identify relevant manifestations. Further, the location of each manifestation can also be visualized for visual interpretability. Our proposed neural model achieved a test AUC of 0.992 on LIDC public dataset and a test AUC of 0.923 on our in-house dataset. Moreover, our experimental results proved that by incorporating manifestation identification tasks into the multi-task model, the accuracy of the malignancy classification can also be improved. This multi-task explainable model may provide a scheme for better interaction with the radiologists in a clinical environment.

## I. INTRODUCTION

In the past few years, artificial intelligence (AI) technology, especially deep-learning (DL), has swept the entire industry and academia, and also has strongly impact radiologic research. A rising number of researches have been published using AI techniques to tackle challenging medical imaging problems [1], and it has been reported frequently that AI models outperformed human clinicians [2-4]. Consequently, more and more AI products emerged in the clinical environment [5] and actually influenced clinicians in daily diagnosis work.

One important reason that DL is gradually replacing traditional machine-learning techniques is that data-driven DL takes full advantages of big-data. It can automatically design the feature space without handcrafted feature engineering. However, lack of model transparency and interpretability is a major obstacle for its large-scale usage in high-stake decision-making, such as healthcare and criminal justice. It is crucial to develop reliable and explainable AI models for clinical practice [6-8]. As a second reader, AI models should provide reliable diagnostic results for radiologists. Here, *reliable diagnostic results* not only refer to accuracy but also explainability. "Opening the black-box" has always been one of the research hotspots.

Two major routines on comprehending AI model focus on "why" and "where", respectively. "Why" approach attempts to give the reasons for the diagnostic results, while "Where" approach tries to visualize the critical region where the AI is looking at.

Ribeiro *et al.* proposed a "LIME" method [9] trying to explain AI model by identifying useful portions in input data, which has been successfully applied in natural language processing (NLP). Same idea has been brought into computer vision. Ghorbani *et al.* presented concept-based explanation approach [10] which attempts to extract human-understandable patches from the input images. Instance-based approaches [11-13] intended to explore the underlying relationship between the training data and a given test data by identifying similar instance from training cohort.

Different from aforementioned approaches which attempt to explain how AI models come to the predictions, class activation mapping (CAM) family [14-16] generates saliency maps to visualize the specific region activated in the AI model. CAM-based approaches have been widely used in many healthcare-related works, such as COVID-19 classification [17], skin lesion classification [18, 19], prediction of prostate cancer extracapsular extension [3], prediction of lymph node status in early-stage breast cancer [20].

Lung cancer has the highest mortality rate of deadly cancers around the world [21]. Computer-aided diagnosis (CAD) on pulmonary nodules have been developing for a quite long time [22-25]. CAD system for pulmonary nodules mainly focus on

---

This paragraph of the first footnote will contain the date on which you submitted your paper for review, which is populated by IEEE. Part of this work is sponsored by Shanghai Pujiang Program (Grant No. 2020PJD016) and China Postdoctoral Science Foundation (Grant No. 2021M691038). *(Corresponding author: Guang Yang).* Chenglong Wang and Yun Liu made equal contributions.

C. Wang, Y. Liu, C. Zhang, Y. Wang and G. Yang are with the Shanghai Key Laboratory of Magnetic Resonance, East China Normal University, Shanghai 200062, China (e-mail: clwang@phy.ecnu.edu.cn; yliumri@gmail.com; cxzhang@phy.ecnu.edu.cn; ydwang@phy.ecnu.edu.cn; gyang@phy.ecnu.edu.cn).

F. Wang and M. Yuan are with Department of Radiology, the First Affiliated Hospital of Nanjing Medical University, Nanjing, Jiangsu Province 210029, China (e-mail: wangfen0102li@163.com; yuanmeijiangsu@163.com).

This article has supplementary downloadable material available at http://ieeexplore.ieee.org, provided by the authors.

Source code of this work will be released soon.

Color versions of one or more of the figures in this article are available online at http://ieeexplore.ieee.org

32
> REPLACE THIS LINE WITH YOUR MANUSCRIPT ID NUMBER (DOUBLE-CLICK HERE TO EDIT) <*detection* [26, 27] and *classification* tasks [28]. However, most of studies emphasize the accuracy and payed no attention to interpretability. Recently, researchers began to realize the importance of interpretability. Several works have developed explainable pulmonary nodules diagnostic AI model by giving predicted clinical manifestations [29, 30], such as calcification, sphericity and subtlety *etc., while other* researchers attempted to visualize the critical area where the AI model is looking at [31, 32].

In this work, we proposed a new explainable AI model which can diagnose pulmonary nodules with high accuracy and provide both "why" and "where" to radiologists. Different from conventional diagnosis system, which only gives one definite diagnosis results, our approach will provide not only predictive diagnosis results, but also relevant manifestations, such as subtlety, calcification, texture, sphericity and margin features, as well as the corresponding saliency maps for localization.

This work makes several contributions to the current literature:

1) We proposed a multi-task explainable deep-learning model for pulmonary nodule diagnosis. Two tasks are designed: the main task aims to learn the main diagnosis task such as benign and malignant, and the secondary task is designed to learn the corresponding manifestations to provide interpretability. Two tasks are trained simultaneously in a supervised fashion to make full use of multi-task learning.

2) We incorporated anatomical attention module into our classification framework to explicitly provide specific spatial attention for the model. Experimental results indicated that explicit attention was more effective than self-attention mechanism on limited data such as our pulmonary nodules. Furthermore, we used automatic nodule segmentation to provide the attention to avoid manual annotations in test phase.

3) Our proposed architecture can be easily used in different scenarios. We evaluated the architecture with two pulmonary nodule datasets with different main and secondary tasks and achieved state-of-the-art performance in both datasets.

4) This study provides important insights into the positive effects of using manifestations for pulmonary nodule diagnosis task. Manifestations can not only provide interpretability for neural model, but also can contribute to an improved accuracy of the diagnostic model.

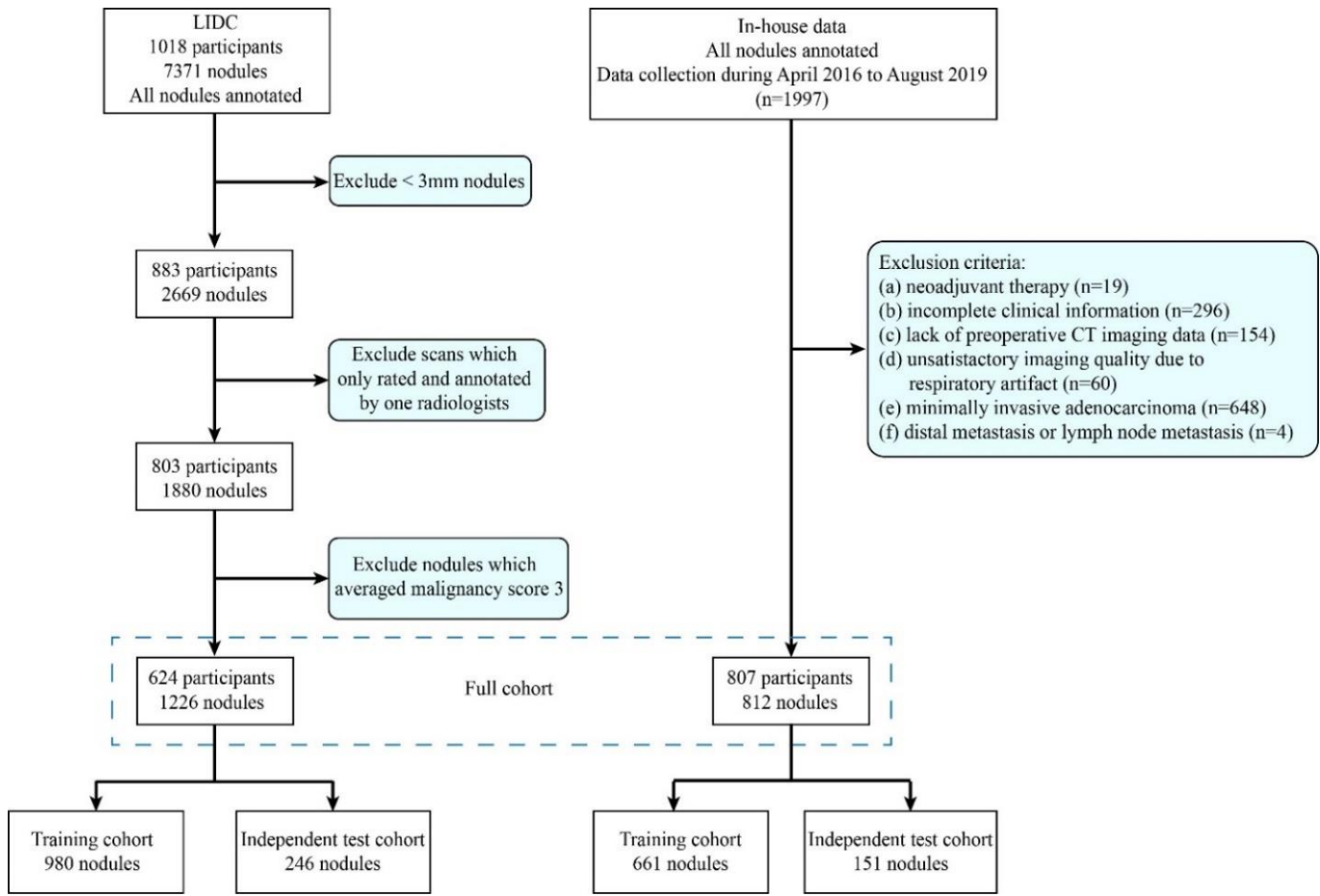

Fig. 1. Enrollment criteria for LIDC dataset and in-house dataset.

**Related works**

Most related to our work is the model of Shen *et al.* [29], which proposed an interpretable deep hierarchical semantic convolutional neural network (HSCNN) for pulmonary nodule classification. HSCNN produces two levels of output, one is high-level prediction of nodule malignancy and the other is low-level semantic features, *i.e.* manifestations. Validation AUC of 0.856 on LIDC dataset was achieved. Based on HSCNN, Lin *et al.* [33] presented an interpretable end-to-end computer-aided detection and diagnosis tool which contains a nodule detector and a nodule malignancy classifier. Liu *et al.* [30] designed a multi-task Siamese network with a margin ranking loss to explore the internal relationship between sub-tasks and achieved a Validation AUC of 0.979 on LIDC dataset. While all aforementioned approaches tried to give manifestations as interpretability for the neural model, our model goes further to provide visual interpretability and demonstrates the state-of-the-art diagnostic accuracy.

Gu *et al.* [34] proposed a visually interpretable network (VINet) which generated visual interpretations while making accurate diagnoses. The main idea of the VINet is to learn an importance map which can contribute to a better classification performance. The advantage of VINet is that it is learned in a target-oriented fashion, which means no manually labeled manifestations are needed for supervised training. However, like CAM approaches, the importance map as visual interpretations only provide a kind of post-hoc explainability. Choi *et al.* [35] proposed an interpretable spiculation feature computed by using the area distortion metric from spherical conformal parameterization. However, the computation of spiculation feature depend heavily on the nodule segmentation, and like many hand-craft features, it faces the generalization problem.

## II. MATERIALS

We used two datasets of thoracic CT images with manually annotated nodule masks, diagnostic labels and manifestations in this study, as described below.

(1) LIDC/IDRI (Lung Image Database Consortium / Image Database Resource Initiative) dataset is an open dataset collected retrospectively from seven academic centers [36]. LIDC dataset was prepared by using officially recommended toolkit "pylidc" [37] to convert data and annotations. The consensual annotations are generated by computing 50% consensus consolidation of all annotation contours. The main task for LIDC is to classify benign and malignant nodules. We used the average scores of malignancy as final diagnosis for all nodules. Nodules with an average malignancy higher than 3 were marked as positive, otherwise marked as negative. Those nodules with an average score equal to 3 were considered indeterminate. We divide the dataset into 2 cohorts: a training cohort and an independent internal test cohort. The inclusion criteria for LIDC were as follows: 1) having diameter of nodule larger than 3mm; 2) rated by at least two radiologists; 3) excluding indeterminate nodules. Finally, we obtained 1226 pulmonary nodules. Detailed enrollment criteria are described in Fig. 1. Detailed characteristics of dataset and sample distributions are listed in **Supplementary (a)**. Besides, LIDC dataset provides five diagnostically relevant manifestations, i.e., subtlety, calcification, texture, sphericity and margin.

(2) Our in-house data is retrospectively collected from The First Affiliated Hospital of Nanjing Medical University, containing 816 nodules from 816 participants. We used this data to differentiate invasive lung cancer with micropapillary or solid pattern (MPL/SOL). The retrospective use of the data was approved by local institutional ethics review board, and the requirement for informed consent was waived. Detailed inclusion and exclusion criteria are also shown in Fig. 1.

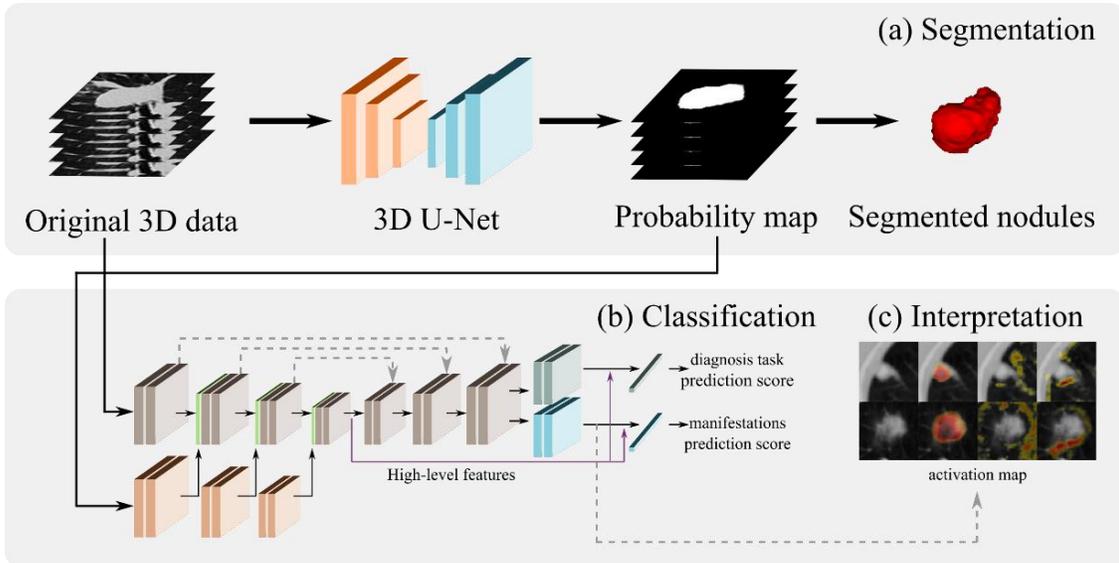

Fig. 2. Overview of proposed approach. The whole framework consists of 3 parts: nodule segmentation, classification, and interpretation. (a) Segment nodules on CT images by using 3D nnU-Net and get the probability map of each nodule. (b) Classify the lung nodules and manifestations simultaneously by using our proposed model. (c) Use the activation map from SAM module to identify where the network is looking at to visualize the manifestations.

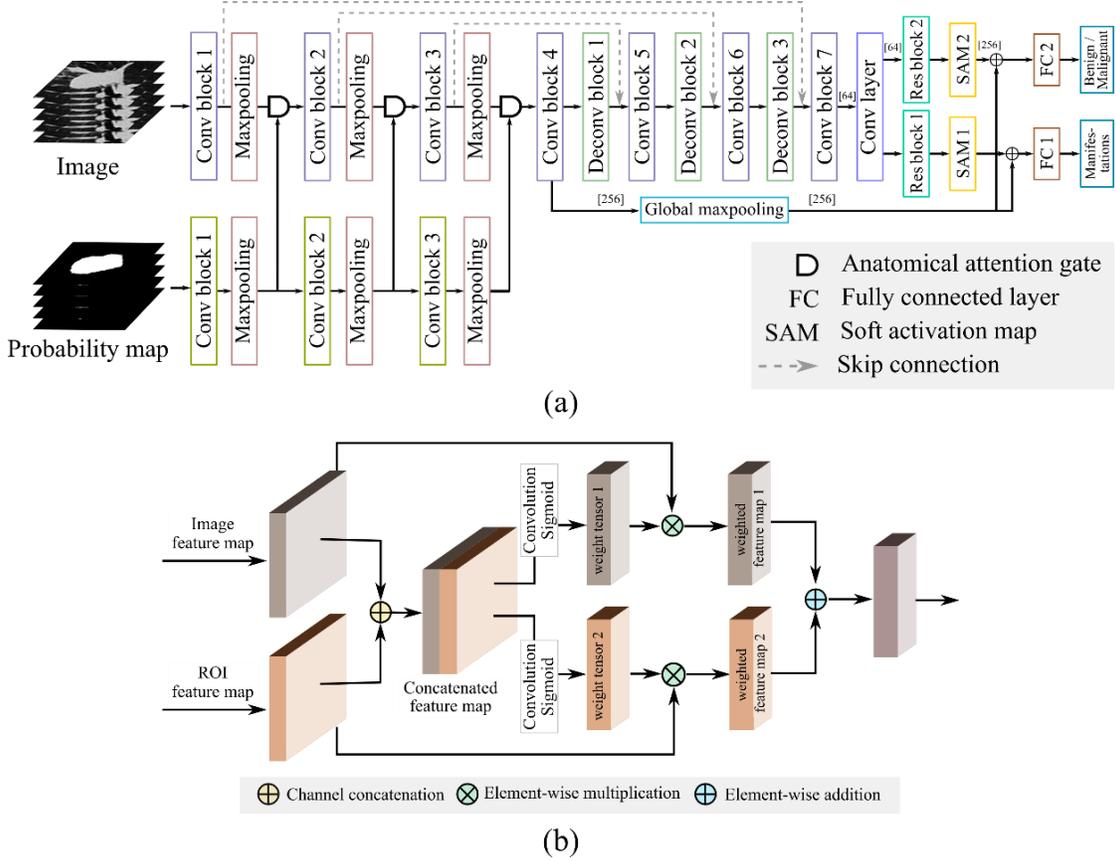

Fig. 3. (a) Structure of proposed multi-task classification network. Both images and probability maps produced by segmentation are fed into the network. In the upper branch, images are fed into a U-Net based structure. In the lower branch, probability maps are passing through the same convolution blocks and max-pooling layer with images. After each max-pooling, the ROI feature maps and image feature maps are fed into AAG module. Finally, the output of U-Net is fed into two separate residual blocks, each followed by corresponding FC layer to get classification results. (b) Anatomical attention gate module. Image feature maps and ROI feature maps are channel-wisely concatenated. The concatenated feature maps are then fed into two separate convolutional layers followed by Sigmoid activations. The generated feature maps will be used for spatial attention.

For the in-house data, we also have 4 manifestations evaluated by radiologists, *i.e.*, lobulation, speculation, relation to the bronchus (RB), relation to the vessel (RV). Presence of lobulation and speculation are defined as positive. RB and RV are defined as positive for lesions with bronchus/vessel passing through. Pathological diagnoses are taken as gold standard for the main task. Detailed characteristics and manifestation descriptions are listed in **Supplementary (a)**.

### III. METHODS

*A. Architecture of neural network*

The whole framework of this study is illustrated in Fig. 2. We modified 2D HESAM [32] to a 3D architecture to be used as our backbone network. HESAM architecture combines both high-level features and soft activation map (SAM) to gain fine-grained attention regions, and achieved the state-of-the-art performance in pulmonary nodule classification. Compared to the coarse-grained activation map provided by conventional CAM module, the fine-grained activation map generated by SAM module can effectively visualize the attention regions of small structures, which is more suitable for pulmonary nodule classification.

Multi-task learning has been successfully applied in predicting pulmonary nodules manifestations [29, 30]. In this work, we also adopted multi-task framework to build our explainable neural network. The architecture of our neural network is illustrated in Fig. 3(a). Two classification heads are constructed for each task. One head is for main diagnosis task, such as benign/malignant classification, the other is designed for multi-label manifestation identification. The manifestations can be provided to radiologists for better understanding the diagnosis of the neural network. We extracted diagnosis and manifestations features by two separate residual blocks in order to extract features that are most relevant to the corresponding task.

Following the original HESAM design, we summed the features generated by SAM module and the high-level features generated by global max-pooling (GMP). The summed features are then fed into two separate fully-connected (FC) layers for diagnosis and manifestation identification, respectively. The activation map from SAM told us "where" the network focuses on, and the output from FC layer hinted us "why" the neural network gives those diagnosis by indicating present manifestations. Furthermore, since the manifestations used in the study are believed to be relevant to diagnosis, by learning manifestation-related features, the neural network can learn more valuable diagnostic representation of pulmonary nodules. To guide the network to focus on the pulmonary nodule, we adopted anatomical attention gate (AAG) module proposed by Sun *et al.* [38]. The original motivation of AAG module was to use an atlas of labeled images to provide information about the anatomical structure of the brain to improve the segmentation accuracy. In this work, we used AAG module to merge ROI information of nodules into main classification branch. We used probability maps produced by segmentation to provide anatomical information, which provides an attention constraint "softer" than binary ROI masks. The detailed structure of AAG module is illustrated in Fig. 3(b).

*B. Implementation Details*

We used a 3D version of HESAM network [32] as our backbone for classification. U-Net like architecture was used for feature extraction. High-level features were obtained from the bottom convolutional layer followed by GMP. The features from the final convolutional layer of the U-Net structure are fed into two separate residual blocks. Each residual block contains two basic convolutional blocks followed by a SAM module. Finally, the summation of the output features of SAM module and the high-level features is fed into two FC layers, one for pulmonary nodule manifestation identification and the other for benign/malignant classification.

As for the anatomical attention branch, three convolutional layers followed by batch normalization, ReLU activation and max-pooling layers are used to extract anatomical features. The stride of max-pooling layer is set to 2. The features down-sampled by max-pooling were then fed into AAG modules.

In the training phase, a batch of image patches of the size of $64 \times 64 \times 32$ were fed into neural network. Since the manifestation identification converged slower than the main task, we firstly trained the manifestation identification head and froze the main task head. After the manifestations were sufficiently learned, we unfroze the main task head and trained the two heads simultaneously until all tasks converged. The best deep-learning model was chosen based on evaluation metric of the validation dataset. We used weighted summation of validation metrics of all tasks for model selection: $\text{argmax}_\theta \mathcal{V}_D + \sum_i^K w_i \mathcal{V}_{M_i}$, where $\theta$ denotes the network parameters, $\mathcal{V}_D$ and $\mathcal{V}_{M_i}$ ($i \in K$) denote validation metrics for diagnosis and $K$ manifestations, respectively. $w_i$ denotes weights for each manifestation. We set $w_i = 1/k$ in all experiments.

## IV. EXPERIMENTS

To evaluate the generalization and robustness of our method, we validate it on both LIDC and in-house data and achieved state-of-the-art performance on both datasets. In this section, we will describe details of experiment design including data preprocessing in subsection *A*. Image annotation and preprocessing, experimental settings for two datasets in subsection *B*. Experimental Settings.

*A. Image annotation and preprocessing*

LIDC dataset employed 4 radiologists to review and annotate cases independently. For each case, we used consensual annotations as final annotations for nodules. For our in-house data, all cases were annotated by a radiologist with 3 years' experience in thoracic imaging and reviewed by a senior radiologist with 12 years' experience in thoracic imaging. First, we resampled all cases to 0.7mm×0.7mm×1.25mm for LIDC dataset and 0.7mm×0.7mm×1.5mm for the in-house dataset, which is the median spacing of the training cohort. Then we clipped the intensity values to the [0.5, 99.5] percentiles and used min-max to normalize the intensity. Finally, image patches of the size of 64×64×32 pixels centered at nodules were extracted.

*B. Experimental Settings*

*1) Segmentation Experimental Settings*
For pulmonary nodule segmentation, we used nnU-Net [39] with the weighted sum of Dice and cross-entropy as loss function:

$$L_{total} = L_{Dice} + wL_{CE} \qquad (1)$$

where $L_{total}$ is total loss, $L_{Dice}$ represents Dice loss and $L_{CE}$ represents cross-entropy loss [40] with weight $w$ which was set to 1 in our experiments. We used stochastic gradient decent (SGD) [41] as our optimizer, with an initial learning rate of 0.01, momentum of 0.99 and weight decay of $3 \times 10^{-5}$. Mini-batch size was 14 and max epoch num was 200. Five-fold cross-validation was conducted in training and the five trained models from all folds were subsequently used as an ensemble for segmentation.

*2) Classification Experimental Settings*
For classification, we used weighted sum of binary cross-entropy loss as loss function:

$$L = y \log \hat{y} + (1-y) \log(1-\hat{y}) \qquad (2)$$
$$L_{all} = L_D + \sum_i^K w_i L_{M_i} \qquad (3)$$

where $L$ is the binary cross-entropy, $\hat{y}$ are labels and $y$ represents the predicted scores. $L_{all}$ denotes the final loss. $L_D$ represents diagnosis loss. $L_{M_i}$ represents the loss for the *i*-th manifestation, and $w_i$ denotes the corresponding weight, which was set to the inverse of the manifestation classification AUC values determined in the first training phase, according to a simple task prioritization scheme [42]. In our experiments, for LIDC dataset, the first-phase AUC values were approximately 0.7, 0.9, 0.9, 0.7, and 0.8 for subtlety, calcification, texture, sphericity and margin manifestations, respectively. For our in-house dataset, the first-phase AUC values were approximately 0.8, 0.85, 0.85, and 0.75 for lobulation, speculation, RB, and RV manifestations, respectively.

We used stochastic gradient decent (SGD) as optimizer, with an initial learning rate of 0.01, which was reduced to 1/10 when the metric did not improve with a patience of 15. Mini-batch size was set to 10. Max epoch number was set to 500 and early stopping with a patience of 30 was used.

Table 1. Comparison of the diagnosis performance of proposed model with typical classification models and previously reported results.

| Dataset | Models | AUC [CI] | Acc | Sen | Prec | F1 |
|---|---|---|---|---|---|---|
| LIDC | VGG | 0.980 [0.885-0.972] | 0.939 | 0.923 | 0.750 | 0.828 |
| | ResNet | 0.974 [0.854-0.963] | 0.943 | 0.872 | **0.791** | 0.829 |
| | DenseNet | 0.960 [0.845-0.929] | 0.854 | 0.949 | 0.521 | 0.673 |
| | HSCNN [29] | 0.856 [-] | 0.842 | 0.705 | - | - |
| | MC-CNN [43] | 0.930 [-] | 0.871 | 0.770 | - | - |
| | MTMR-Net [30] | 0.979 [-] | 0.935 | 0.930 | - | - |
| | Our proposed | **0.992** [0.984-0.998] | **0.955** | **1.000** | 0.780 | **0.876** |
| In-house | VGG | 0.852 [0.708-0.846] | 0.808 | 0.673 | 0.771 | 0.718 |
| | ResNet | 0.824 [0.722-0.856] | 0.795 | 0.782 | 0.694 | 0.735 |
| | DenseNet | 0.867 [0.762-0.885] | 0.821 | 0.836 | 0.719 | 0.773 |
| | Our proposed | **0.923** [0.873-0.965] | **0.901** | **0.909** | **0.833** | **0.870** |

* Results of HSCNN, MC-CNN, and MTMR-Net come from the literature. CI represents for confidence interval.

Table 2. Comparison of the performance of manifestation identification of the proposed model and previously reported results.

| | Models | Subtlety | | Calcification | | Texture | | Margin | | Sphericity | |
|---|---|---|---|---|---|---|---|---|---|---|---|
| | | AUC | Acc | AUC | Acc | AUC | Acc | AUC | Acc | AUC | Acc |
| LIDC | HSCNN [29] | 0.803 | 0.719 | 0.930 | 0.908 | 0.850 | 0.834 | 0.776 | 0.725 | 0.568 | 0.552 |
| | HESAM + multitask + AAG | **0.832** | **0.736** | **0.967** | **0.923** | **0.947** | **0.898** | **0.840** | **0.634** | **0.882** | **0.805** |
| | | Lobulation | | Speculation | | RB | | RV | | | |
| In-house | | AUC | Acc | AUC | Acc | AUC | Acc | AUC | Acc | | |
| | HESAM + multitask + AAG | **0.817** | **0.781** | 0.822 | **0.801** | 0.858 | 0.781 | 0.723 | 0.649 | | |

Table 3. Ablation study of the diagnostic performance on both LIDC and in-house dataset.

| Dataset | Models | AUC [CI] | Acc | Sen | Prec | F1 |
|---|---|---|---|---|---|---|
| LIDC | HESAM [32] | 0.978 [0.788-0.928] | 0.939 | 0.744 | 0.853 | 0.795 |
| | HESAM + multitask | 0.986 [0.971-0.997] | 0.943 | 0.949 | 0.755 | 0.841 |
| | HESAM + multitask + CBAM | 0.988 [0.974-0.998] | 0.971 | 0.923 | **0.900** | **0.911** |
| | HESAM + multitask + AAG | **0.992** [0.984-0.998] | 0.955 | **1.000** | 0.780 | 0.876 |
| In-house | HESAM [32] | 0.873 [0.809-0.929] | 0.828 | 0.819 | 0.738 | 0.776 |
| | HESAM + multitask | 0.887 [0.827-0.941] | 0.861 | 0.891 | 0.766 | 0.824 |
| | HESAM + multitask + CBAM | 0.871 [0.803-0.932] | 0.854 | 0.727 | 0.851 | 0.784 |
| | HESAM + multitask + AAG | **0.923** [0.873-0.965] | **0.901** | **0.909** | **0.833** | **0.870** |

* Multitask denotes using multi-task framework, CBAM[44] denotes convolution block attention module, AAG denotes using anatomical attention gate module. CI represents for confidence interval.

*C. Results*

In this section, we will demonstrate our experimental results of both LIDC dataset and our in-house data. Quantitative analysis was evaluated by using the area under the ROC curve (AUC), accuracy (*Acc*), sensitivity (*Sen*) also known as recall, Precision (*Prec*) and F1 score (*F1*). All of these statistical values were estimated by using scikit-learn package [45]. Quantitative segmentation and classification results are described in subsection *C.1* and *C.2*. To demonstrate the interpretability of our neural model, we visualized network activation maps corresponding to each manifestation shown in subsection *C.3*.

Table 4. Ablation study of the manifestation identification.

| | Models | Subtlety | | Calcification | | Texture | | Margin | | Sphericity | |
|---|---|---|---|---|---|---|---|---|---|---|---|
| | | AUC | Acc | AUC | Acc | AUC | Acc | AUC | Acc | AUC | Acc |
| LIDC | HSCNN [29] | 0.803 | 0.719 | 0.930 | 0.908 | 0.850 | 0.834 | 0.776 | 0.725 | 0.568 | 0.552 |
| | HESAM + multitask | 0.680 | 0.500 | 0.828 | 0.711 | 0.594 | 0.512 | 0.660 | 0.800 | 0.794 | 0.724 |
| | HESAM + multitask + CBAM | 0.692 | 0.663 | 0.778 | 0.703 | 0.697 | 0.459 | 0.699 | 0.748 | 0.758 | 0.699 |
| | HESAM + multitask + AAG | **0.832** | **0.736** | **0.967** | **0.923** | **0.947** | **0.898** | **0.840** | 0.634 | **0.882** | **0.805** |

| | | Lobulation | | Speculation | | RB | | RV | |
|---|---|---|---|---|---|---|---|---|---|
| | | AUC | Acc | AUC | Acc | AUC | Acc | AUC | Acc |
| In-house | HESAM + multitask | 0.797 | 0.775 | **0.852** | 0.755 | 0.825 | 0.781 | 0.637 | **0.656** |
| | HESAM + multitask + CBAM | 0.728 | 0.715 | 0.832 | 0.762 | 0.809 | 0.722 | 0.638 | 0.570 |
| | HESAM + multitask + AAG | **0.817** | **0.781** | 0.822 | **0.801** | **0.858** | **0.781** | **0.723** | 0.649 |

**\*** Multitask denotes using multi-task framework, CBAM[44] denotes convolution block attention module, AAG denotes using anatomical attention gate module.

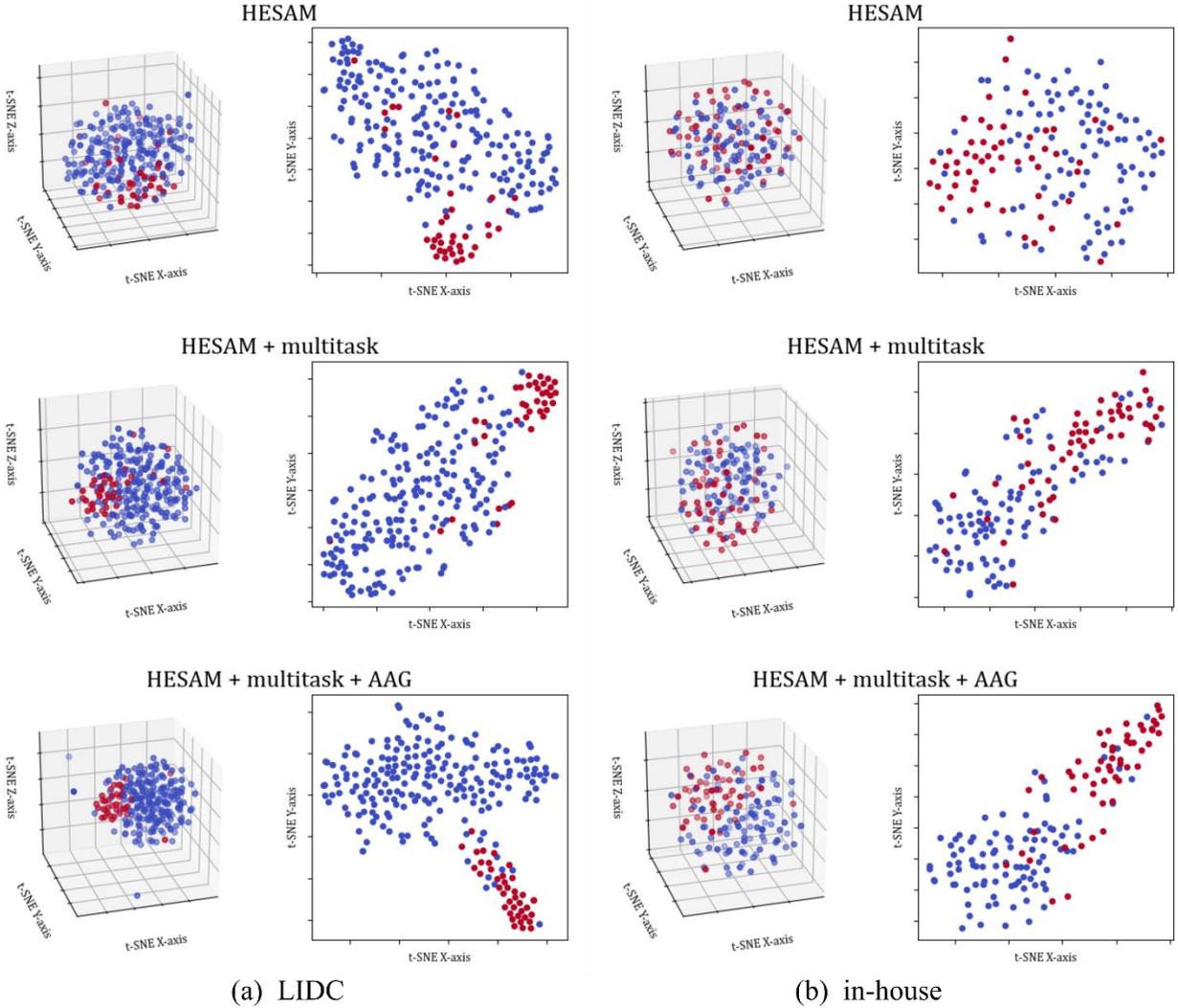

(a) LIDC  (b) in-house

Fig. 4. Visualization of dimension-reduced features on both LIDC and in-house dataset. Dimension reduction is performed by t-SNE [46]. Three- and two-dimensional features are shown on the left and right respectively. Red represents positive samples.

*1) Segmentation performance*

The average Dice of the LIDC and in-house test cohorts were 0.818 and 0.867, respectively. Since the average Dice of manual segmentation between radiologists in LIDC test cohort was 0.73, it can be concluded that the automatic pulmonary nodule segmentation has achieved acceptable accuracy.

*2) Classification performance*

We compared our method with some widely used classification networks and related pulmonary nodules classification works.

First, we systematically performed quantitative analysis of the performance of the main diagnosis task and the auxiliary manifestation identification task on both LIDC and our in-house dataset, as shown in Table 1 and Table 2. From these tables, we can see that the proposed network outperformed other modern CNN networks and related works on LIDC dataset. On our in-house, our model outperformed other models in all metrics. Identification of manifestations also achieved reasonable high performance compared with previous works.

Ablation experiments were performed to quantitatively demonstrate the contributions of the multi-task framework and AAG module, for both diagnosis and manifestation identification, as shown in Table 3 and Table 4, respectively. It can be seen from these two tables that the model integrating multi-task framework and AAG achieved the best performance. It can be seen that the AAG module significantly improved the performance of both diagnosis and manifestations identification.

Furthermore, feature dimension reduction was also performed to compare the feature distributions of deep latent representations. We applied t-SNE visualization to the input features of the final fully connected layer of neural models. The experimental results are shown in Fig. 4, and more detailed comparisons between different neural networks are illustrated in **Supplementary (b)**. As shown in Fig. 4, our proposed model can generate better discriminative latent representation compared with baseline model.

*3) Interpretability*

In this section, we investigate the visual interpretability of manifestations. We used the activation map generated by SAM module to visualize location of manifestations. Similar to VINet [34], SAM was trained in a target-oriented fashion. Compared with CAM-based approaches which used a trained model to visualize, SAM can reflect the activated region more accurately.

Activation maps of cases with typical manifestations from two datasets are shown in Fig. 5 and Fig. 6, respectively. The activation maps were normalized to 0-1. For (a), (b) and (e) in Fig. 5, values below 0.25 are not displayed for better visualization. For each case, the original image patch, image patch with overlapped activation map and corresponding boundaries are shown. For "margin" manifestation, axial, sagittal and coronal views are shown in Fig. 5(e).

It can be observed that our model clearly highlights the region of calcification shown in Fig. 5(a), and focuses on ground glass texture of nodules shown in Fig. 5(b). In Fig. 5(c), neural model is looking at whole background region to identify whether the nodule is easy to be found. In Fig. 5 (e), the network mainly focuses on the margin of the nodules to decide whether the nodule has a well-defined margin.

As demonstrated in Fig. 6, our model precisely points out the lobulation and speculation manifestations shown in Fig. 6(a) and (b). For RB and RV manifestations, our model also attempts to find bronchus and vessel-like structures shown in Fig. 6(c) and (d). The hot areas on activation maps are also outlined for better visualization.

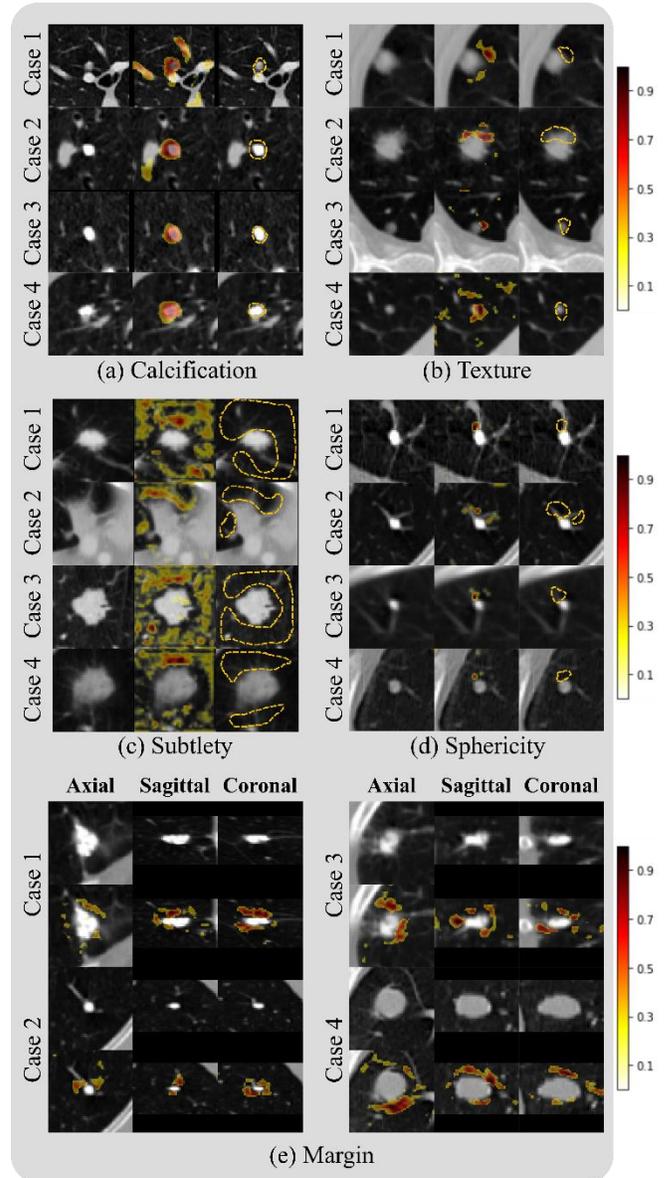

(a) Calcification  (b) Texture

(c) Subtlety  (d) Sphericity

(e) Margin

**Fig. 5.** Activation maps generated by our proposed model on LIDC dataset. (a), (b), (c), and (d) represent cases with calcification, texture, subtlety and sphericity manifestation respectively where left, middle and right columns for each case denotes original image patch, patch with overlapped activation map and patch with map boundary. (e) represents cases with margin manifestation where first and second rows for each case denotes original image patch and patch with overlapped activation map, respectively. Three columns in (3) denotes axial, sagittal and coronal views respectively.

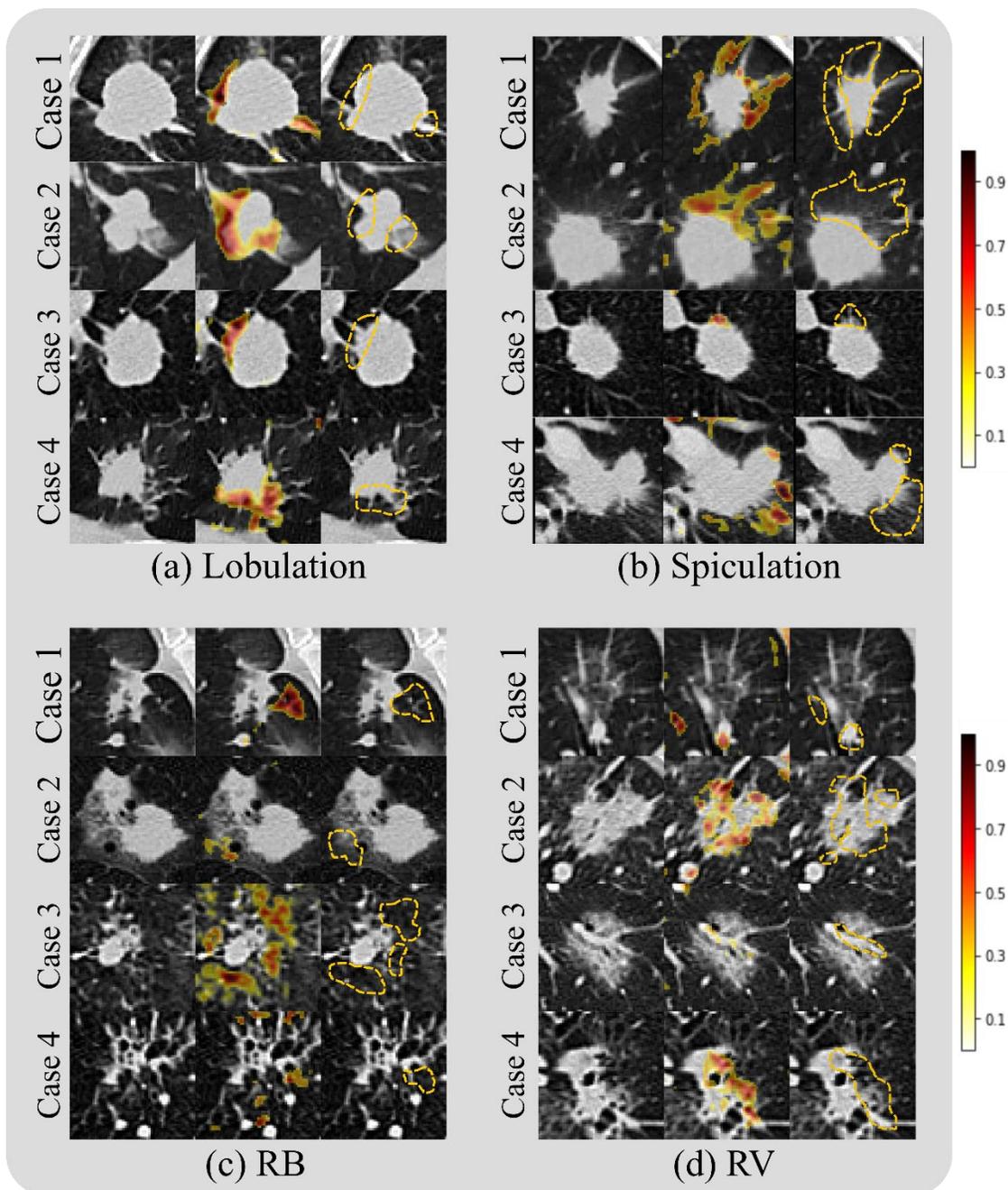

**Fig. 6.** Activation maps generated by our proposed method on in-house dataset. (a), (b), (c) and (d) demonstrate typical cases with lobulation, speculation, RB (relation to bronchus), and RV (relation to the vessel) manifestations. Left, middle and right columns for each case denotes original image patch, patch with overlapped activation map and patch with map boundary, respectively. Boundaries are drawn manually to illustrate the activation regions more clearly.

## V. DISCUSSION

As deep-learning technology sweeps across all fields, more and more AI products are used in clinical settings, and begin to influence clinicians in their daily work. While most deep learning studies in medical imaging focus on improving classification accuracy, the importance of interpretability should not be ignored. Lack of model transparency and interpretability can greatly impede the large-scale application of AI in high-stakes decision-making fields, such as healthcare and criminal justice [47].

In this work, we presented a multi-task explainable deep-learning model for pulmonary nodule diagnosis to fulfil the requirement of high diagnostic accuracy and clinical interpretability. We designed two classification head for malignancy classification and manifestation identification, respectively. Compared with related works, the proposed method can not only produce diagnostic conclusions and corresponding manifestations, but also give a fine-grained visual interpretation map as supportive indicators.

The primary contribution of this work is that we proposed a multi-task explainable deep-learning model for pulmonary nodule diagnosis. Compared to previous works [29, 30, 33, 34], our work achieved state-of-the-art diagnostic performance, and obtained high classification accuracy of manifestations. From Table 3, it can be seen that by introducing manifestation identification task, the performance of malignancy classification has also been significantly improved. It confirmed our hypothesis that manifestation identification can help network to learn a more accurate diagnostic representation, which are also consistent with the diagnosis process of radiologists. HSCNN [29] also used multi-task network for manifestation identification. Compared to HSCNN, we not only greatly improved the diagnostic accuracy, but also provided fine-grained visual interpretation activation maps for relevant manifestations. Most previous explainable works on lung cancer [33, 34, 48] only gave activation maps of malignancy classification, but ignored the diagnostic manifestations. Detailed activation maps of manifestations can improve radiologists' confidence to assess the diagnostic results. We adopted multi-task learning to implement simultaneous training of diagnosis and manifestation identification, using HESAM network as backbone to provide fine-grained visual interpretability, and using AAG attention module to further improve classification performance. By incorporating these approaches, we presented a model with higher pulmonary nodule diagnosis accuracy and better interpretability. The usability of our approach was demonstrated with two datasets with different diagnosis problem and different sets of manifestations.

Explicit attention mechanism like AAG module was reported to be more effective than self-attention module such as CBAM module [44], especially for medical images with larger data dimension but smaller data size. Our quantitative comparisons have proved this, as demonstrated in Table 3. The disadvantage of AAG is that it needs additional information to provide explicit attention. In this work, we adopted automated segmentation to provide attention to decrease manual labor and increase annotation consistency.

In this work, we not only give predictive scores of manifestations but also provide visual interpretation of each manifestation. As shown in Fig. 5 and Fig. 6, some visual maps are very intuitive, showing why and where manifestations such as calcification, texture, subtlety, margin, lobulation and speculation were identified. However, the visual maps of sphericity, RB and RV manifestations are not as intuitive. From Fig. 5(d), it can be observed that the network seems to focus on irregular region of nodules. As shown in Fig. 6(c) and (d), networks seemingly failed to find out bronchus and vessels. Due to the limited number of positive cases of RB (185 positives out of 661 training cohort) and RV (155 positives out of 661 training cohort) manifestations, increasing data size may improve the performances. Another possible solution is to explicitly input bronchi and vessels ROI information as auxiliary anatomical attentions. Simultaneous segmentation of nodule, bronchi and vessels could be one of directions of future works.

A considerable amount of studies has been done to deal with the training issue between subtasks in multi-task learning [49]. In this work, we simply apply two-step cascade training strategy to deal with the different converge speed between main and secondary tasks. This naive optimization strategy did not consider the task balancing problem. More reasonable optimization strategy should be considered in the future work to explore a better optimization.

## VI. CONCLUSION

In this work, by combining pulmonary nodule diagnosis and manifestation identification in a multi-task deep-learning model, we achieved a higher diagnostic accuracy and better interpretability at the same time. Different with previous related works, the proposed model can not only indicate the present of related manifestations but can also provide visual indications showing where to look for each manifestation. We believe this can help radiologists better understand the diagnosis of the model. Furthermore, through the quantitative analysis, it was demonstrated that incorporating relevant manifestations into the model can help improve the diagnosis performance, implying that the pursuit for interpretability does not necessarily conflict with the target of high-precision diagnosis. Thus, we hope this study can inspire more studies towards mutual improvement of model performance and interpretability, which will surely facilitate the application of AI in medical imaging.